# Reply to "Rebuttal of "The multicaloric effect in multiferroic materials", [arXiv:1602.04238]"

Melvin M. Vopson[*]

Faculty of Science, SEES, University of Portsmouth, Portsmouth PO1 3QL, UK

[*] Formerly known as Vopsaroiu. Contact email: **melvin.vopson@port.ac.uk**

Vopson first published the multicaloric effect in multiferroics in 2012 [M.M. Vopson, Solid State Communications 152, 2067–2070 (2012)]. However, a closer examination of the multicaloric effect and its derivation leads to a contradiction, in which the predicted changes in one of the order phase at a constant applied field are due to the excitation by the same field. This apparent paradox has triggered the publication of the "Rebuttal of "The multicaloric effect in multiferroic materials", [arXiv:1602.04238]" by Starkov et al. This article is the response to their rebuttal.

## 1. Introduction and historic background

Multiferroics are very interesting materials with multi-functional properties. Following the initial surge in interest in the field of magneto-electric multiferroics in 1950s and 1960s [1-4], the topic has lost somehow interest with the scientific community. Recently, the interest in magneto-electric multiferroics grew substantially [5] because of the realization of their potential for technological applications. Wood and Austin summarized many possible applications of multiferroics in an article published in 1973 [6]. Vopson published a recent comprehensive review article detailing possible modern applications of multiferroics in 2015 [7]. One such recently proposed application of multiferroics is their utilization to ultra-efficient solid-state refrigeration via a new effect, the multicaloric effect. Fahler et al. first coined the term "multicaloric" in 2012, although no attempt to provide any experimental or theoretical treatment has been made in their 2012 paper [8]. Later the same year, Vopson proposed, for the first time, the existence of a multicaloric effect [9], which he predicted to occur in multiferroic materials [*M.M. Vopson, The Multicaloric Effect in Multiferroic Materials, Solid State Communications 152 (2012) 2067-2070*]. This is the paper discussed in the current rebuttal article [arXiv:1602.04238]. In his original paper, Vopson showed analytically that coupling in multiferroic materials could substantially enhance the caloric properties of these solids, and he termed this the "Multicaloric Effect in Multiferroics". Vopson also published, a year later, an extension of the 2012 model [10]: *Theory of Giant Caloric Effect in Multiferroic Materials, J. Phys. D: Appl. Phys. 46 345304 (2013).*

In 2014, 2 years after the first publication of the Multicaloric Effect by Vopson, A.S. Starkov and I.A. Starkov published the article "Multicaloric Effect in a Solid: New Aspects" [11]. Despite being 2 years late in predicting this effect, the authors "forgot" to make any reference to previously peer review published papers on this effect, including the original article [9] and the follow up published a year later [10]. The same authors continued to publish various articles on the same topic again, ignoring previously published work. Three years after the original publication of the Multicaloric Effect [9], A.S. Starkov and I.A. Starkov published a rebuttal article [arXiv:1602.04238], to which I am very pleased to offer a response. It is important to point out that Starkov et al. first attempted to publish their rebuttal article in Solid State Communications. Naturally, the journal offered the opportunity for a response from Vopson, and a panel of reviewers plus the editorial board accepted the response and rejected outright Starkov's rebuttal article. The response to their rebuttal has been later submitted as a full journal paper, which, after peer review, has been accepted for publication [12]. Here, the response to their rebuttal is mostly based on the peer review article already published [12].

**2. The problem**

The multicaloric effect is defined as the adiabatic temperature change in multiferroics activated by a single excitation (electric, magnetic or elastic) and it is mathematically described by the general equation of the giant multicaloric effect:

$$\Delta T = -\frac{T}{C} \cdot \sum_{i;i\neq j} \int_{x_j} \left[ \left(\frac{\partial X_i}{\partial T}\right)_{x_j} \cdot \frac{\alpha_{ij}}{\chi_i} + \left(\frac{\partial X_j}{\partial T}\right)_{x_i} \right] \cdot dx_j \qquad (1)$$

where: $X_i$ = Magnetization (M), Polarization (P), Volume (V), Strain (ε),… are the independent variables; $x_i$ = magnetic field (H), electric field (E), mechanical stress (σ),… are the generalized forces / fields thermodynamically conjugated to the generalized variables / displacements $X_i$; $\chi_i$ is the generalized susceptibility in the linear approximation $(\partial X_i / \partial x_i) = \chi_i$; $\alpha_{ij}$ is the generalized linear magneto-electric coupling coefficient $\alpha_{ij} = \alpha_{ji} = (\partial X_j / \partial x_i)_{T, xj\neq i} = (\partial X_i / \partial x_j)_{T, xi\neq j}$; T is the operation temperature and it is a constant; C is defined as the heat capacity of the system at the operation temperature, T, also assumed constant, but it is acknowledged that in reality this is a strong approximation as the heat capacity has some non-negligible variation with the applied fields. A full derivation of relation (1) is given in [10], showing that the cross couplings between displacements and fields / forces play an important role in the multicaloric effect. For a finite adiabatic change in the applied external field $x_i$, a variation in temperature $\Delta T$ is produced resulting in the enhancement of the total temperature change $\Delta T$ due to the cross coupling additional terms $\alpha_{ij}/\chi_i \cdot (\partial X_i / \partial T)$. If (1) is applied to the particular case of a multiferroic material containing electric and magnetic order phases, the electrically and magnetically induced multicaloric effects are described by:

$$\Delta T_E = -\frac{T}{C} \cdot \int_{E_i}^{E_f} \left[ \frac{\alpha_e}{\mu_0 \chi^m} \cdot \left(\frac{\partial M}{\partial T}\right)_{H,E} + \left(\frac{\partial P}{\partial T}\right)_{H,E} \right] \cdot dE \qquad (2)$$

$$\Delta T_H = -\frac{T}{C} \cdot \int_{H_i}^{H_f} \left[ \left(\frac{\partial M}{\partial T}\right)_{H,E} + \frac{\alpha_m}{\varepsilon_0 \chi^e} \cdot \left(\frac{\partial P}{\partial T}\right)_{H,E} \right] \cdot dH \qquad (3)$$

However, the derivation of relations (1)-(3) in ref. [9,10] appears to contain an apparent error, which would invalidate the proposed multicaloric effect. Starkov et al. have pointed out this in their rebuttal article [arXiv:1602.04238]. The following discussion is limited to the particular case of a multiferroic material containing electric and magnetic order phases, for which multicaloric effects are described by (2) and (3).
Applying an external E field to a dielectric polar material has the effect of changing its electric polarization. If the material is a multiferroic, the effect of the E field application is to modify its electric polarization and magnetization, according to the magneto-electric coupling effect. The converse magneto-electric effect is also valid, when a magnetic field applied to a multiferroic results in changes of not only magnetization of the material, but also its electric polarization. Thermodynamically, the magneto-electric coupling coefficient results immediately from Maxwell equations applied to the differential Gibbs free energy of a multiferroic system, without any reference to possible microscopic mechanisms responsible for the coupling [7]. One way of approaching this problem is to regard the magneto-electric effect as the result of fictitious fields induced spontaneously in the material. That is equivalent to saying that the magnetization changes in a multiferroic in response to an applied electric field are due to a spontaneously induced magnetic field inside the material. The converse effect also produces a spontaneous electric field inside the material as a result of the application of a magnetic field. The introduction of the fictitious field is critical for the derivation of relations (2) and (3). Let us now examine this approach for the electrically induced magneto-electric effect, which is mathematically expressed in integral form as:

$$dM = \alpha_e dE \tag{4}$$

where $\alpha_e$ is the electrically induced magneto-electric effect. However, magnetization can be expressed in terms of linear change with a magnetic applied field as:

$$dM = \mu_0 \chi^m dH \tag{5}$$

From the two relations, one could easily deduce the induced magnetic field by an electric applied field in a multiferroic as:

$$dH = \frac{\alpha_e}{\mu_0 \chi^m} dE \tag{6}$$

Relation (6) and its electrical equivalent have been used in [9] to derive the multicaloric effect equations (1) and (2). However, at a closer inspection, as pointed out by Starkov et al., it appears that relation (6) is incorrect. Let us write the differential expression of the magnetization, talking in account the independent variables and constants:

$$dM = \left(\frac{\partial M}{\partial T}\right)_{H,E} dT + \left(\frac{\partial M}{\partial E}\right)_{H,T} dE + \left(\frac{\partial M}{\partial H}\right)_{T,E} dH \tag{7}$$

From (7), the correct expression of relation (4) is written by talking in account that the change in magnetization occurs at constant H and T is:

$$dM_{T,H} = \alpha_e dE \tag{8}$$

where $\alpha_e$ is the electrically induced ME coupling coefficient. From (7), we can also extract the change in magnetization that occurs at constant T and E, which is the equivalent of expression (5):

$$dM_{T,E} = \mu_0 \chi^m dH \tag{9}$$

If we now recombine (8) and (9) to obtain relation (6), we get:

$$dM_{T,E} = dM_{T,H} = \alpha_e dE \tag{10}$$

Relation (10) leads to a contradiction, in which the change in magnetization obtained under constant electric field is related to the change in the electric field, $dM_{T,E} = \alpha_e dE$, which is clearly erroneous, as this would be zero. This fallacy would make relation (6) invalid, indicating that there can be no induced magnetic field by an electric field and vice versa in multiferroics. It would also invalidate the Multicaloric Effect equations (1) - (3) published in [9,10] as claimed by Starkov et al. in their rebuttal article.

**3. The solution**

In classical terms, at constant temperature, stress and other environmental conditions, magnetization can only be modified by the application of a magnetic field. Therefore, the change in magnetization *dM* in multiferroics due to the application of the electric field *dE* can be attributed to a <u>fictitious</u> magnetic applied field, induced by the magneto-electric effect $dH_{me}$. This field must not be confused with *dH* used in relation (7), as this field results from the magneto-electric coupling and it is induced internally by the application of the *E* field. Introducing the <u>fictitious</u> *H* field is very common in other well-established theories. For example, in the molecular field theory of ferromagnetism, Weiss treated ferromagnets as paramagnets by postulating the existence of an additional <u>fictitious</u> magnetic field whose origin is not given. Weiss [13] called this a *molecular field* [P. Weiss, *L'Hypothese du champ Moleculaire et de la Propriete Ferromagnetique*, J. de Phys. **6**, (1907) pp. 661-690]. Neel also used a similar approach when he introduced a <u>fictitious</u> fluctuating magnetic field in order to explain time dependent and relaxation phenomena in magnetic materials. Moreover, introducing a <u>fictitious</u> magnetic field in order to explain the magnetization change due to an applied electric field in multiferroics, is not a new idea. Rado et al. has introduced this idea in his 1961 papers [2,3], which were some of the very first studies of the magneto-electric coupling in $Cr_2O_3$. However, Agyei & Birman wrote one of the best papers detailing and categorizing various conditions of the occurrence of a spontaneous internal E or H field due to the changes in P or M as a result of the application of H or E, respectively [14].

With this in mind, let us re-write relation (7), by taking into account the applied magnetic field and the induced magnetic field due to the magneto-electric effect. This is achieved by replacing *dH* with:

$$dH = dH_{app} - dH_{me} \tag{11}$$

Relation (11) shows that the acting magnetic field on the multiferroic is a combination of the externally applied magnetic field, $H_{app}$ and the induced magneto-electric field, $H_{me}$, where the minus sign indicates that the induced magneto-electric field is opposed to the applied field as dictated by Newton's 3rd law, or its extension to magnetic phenomena, Lenz's law. Using (11), relation (7) should then be written as:

$$dM = \left(\frac{\partial M}{\partial T}\right)_{H,E} dT + \left(\frac{\partial M}{\partial E}\right)_{H,T} dE + \left(\frac{\partial M}{\partial H}\right)_{T,E} dH_{app} - \left(\frac{\partial M}{\partial H}\right)_{T,E} dH_{me} \qquad (12)$$

Electrically induced magneto-electric effect implies that there is no external applied magnetic field, or the magnetic applied field is constant (i.e. change of M due to E applied only). Hence, d$H_{app}$ = 0 and the third term in (12) vanishes. On the other hand, the induced field d$H_{me}$ is the proposed fictitious field that occurs spontaneously due to the application of the *E* field. According to Lenz's law, the strength of the induced field d$H_{me}$ is exactly as large as needed to account for the change in magnetization *M* via the well-known magneto-electric effect due to the application of *E*. This is also the meaning of the negative sign in (11). Therefore, at constant *T* and *H* (or zero applied magnetic field), the change *dM* should be zero as the effect of the applied *E* field on *M* is cancelled out by the occurrence of the induced field. Relation (12) is then:

$$dM_{T,H_{app}=0} = \left(\frac{\partial M}{\partial E}\right)_{H_{app}=0,T} dE - \left(\frac{\partial M}{\partial H}\right)_{T,H_{app}=0} dH_{ME} = 0 \qquad (13)$$

Relation (13) leads to:

$$\left(\frac{\partial M}{\partial E}\right)_{H_{app}=0,T} dE = \left(\frac{\partial M}{\partial H}\right)_{T,H_{app}=0} dH_{ME} \qquad (14)$$

or re-written as:

$$\alpha_e dE = \mu_0 \chi^m dH_{ME} \qquad (15)$$

Finally, from (15) we re-obtain relation (6) of the induced magnetic field due to the application of an electric field in a magneto-electric multiferroic:

$$dH_{ME} = dH = \frac{\alpha_e}{\mu_0 \chi^m} dE \qquad (16)$$

Relation (16) is identical to (6). The same formalism can be easily applied to the converse magneto-electric effect leading to induced internal electric field due to the application of a magnetic field:

$$dE_{ME} = dE = \frac{\alpha_m}{\varepsilon_0 \chi^e} dH \qquad (17)$$

It is important to mention that this is a simplified thermodynamic approach where the vector and tensor components have been neglected. For example the magneto-electric coupling coefficient is, strictly speaking, a second rank tensor with 9 components, $\alpha_{ij}$. In most cases the crystal symmetry and/or experimental geometry reduces significantly the magneto-electric coupling tensor to one or a few dominant components, but the applicability of these relations must be strictly considered in terms of E and H field vector components and their relationship to the sample geometry, crystal symmetry and directions of polarization and magnetization. This is particularly important when an antiferromagnetic phase exists in the multiferroic or when the sample geometry displays strong spatial variations.

**4. Conclusions**

The authors of the rebuttal article had a valid concern, only because they have ignored the statements made in the original Multicaloric Effect manuscript regarding the occurrence of the fictitious E field in response to an applied H field and vice versa. Leaving the thermodynamic treatment aside, let's examine again the physics of the Multicaloric effect proposed by Vopson: The magnetic or electric entropy of a system can be changed by the application / removal of external fields. When only one field, H for example, is applied to a magnetic material adiabatically, the magneto-caloric effect occurs. If the magnetic material is a multiferroic, then the application of H will change not only magnetic entropy, but also electric entropy because of the ME coupling which allows H to change P. This can be mathematically accounted by introducing the fictitious fields as described above. The ME coupling is an undisputed fact. The caloric and entropy change are also undisputed facts. In the original paper [9], the concept of multicaloric effect was first introduced and the original equations describing the effect were derived. This allows anyone to test experimentally the effect as the equations are derived and clearly published. There are indeed very big question marks regarding the Multicaloric Effect. These are not related to the correctness of the approach published by Vopson, but to more serious issues related to the lack of suitable samples, the mismatch of thermal conductivities and heat capacities in multiferroics and the lack of strongly coupled ME effects in which the magnetic and electric phase transition temperatures coexist. These are very big limitations of the Multicaloric Effect proposed by Vopson and attempts to solve some of these issues have been made in another recent article [15].